\documentclass[letterpaper]{article} 
\usepackage{aaai2026}  
\usepackage{times}  
\usepackage{helvet}  
\usepackage{courier}  
\usepackage[hyphens]{url}  
\usepackage{graphicx} 
\usepackage{natbib}  
\usepackage{caption} 
\usepackage{algorithm}
\usepackage{algorithmic}

\usepackage{amsmath}
\usepackage{multirow}
\usepackage{booktabs}
\usepackage{pifont}

\usepackage{newfloat}
\usepackage{listings}
\DeclareCaptionStyle{ruled}{labelfont=normalfont,labelsep=colon,strut=off} 
\floatstyle{ruled}
\newfloat{listing}{tb}{lst}{}
\floatname{listing}{Listing}
\title{An LLM-Based Simulation Framework for
\\Embodied Conversational Agents in Psychological Counseling
}
\author{
    Written by AAAI Press Staff\textsuperscript{\rm 1}\thanks{With help from the AAAI Publications Committee.}\\
    AAAI Style Contributions by Pater Patel Schneider,
    Sunil Issar,\\
    J. Scott Penberthy,
    George Ferguson,
    Hans Guesgen,
    Francisco Cruz\equalcontrib,
    Marc Pujol-Gonzalez\equalcontrib
}

\author{
    Lixiu Wu\equalcontrib\textsuperscript{\rm 1}, 
    Yuanrong Tang\equalcontrib\textsuperscript{\rm 1}, 
    Qisen Pan\textsuperscript{\rm 1}, 
    Xianyang Zhan\textsuperscript{\rm 1}, 
    Yuchen Han\textsuperscript{\rm 1}, 
    Lanxi Xiao\textsuperscript{\rm 1},\\
    Tianhong Wang\textsuperscript{\rm 1}, 
    Chen Zhong\textsuperscript{\rm 1}, 
    Jiangtao Gong\textsuperscript{\rm 1}\thanks{Corresponding author.}
}
\affiliations{
    \textsuperscript{\rm 1}AI Industry Research, Tsinghua University, Beijing, China\\
    wulx1102@gmail.com, 
    tangxtong2022@gmail.com, 
    gongjiangtao@air.tsinghua.edu.cn
}

\usepackage{bibentry}

\begin{document}

\nocopyright

\maketitle

\begin{abstract}
Due to privacy concerns, open dialogue datasets for mental health are primarily generated through human or AI synthesis methods. However, the inherent implicit nature of psychological processes, particularly those of clients, poses challenges to the authenticity and diversity of synthetic data. In this paper, we propose ECAs (short for \underline{E}mbodied \underline{C}onversational \underline{A}gent\underline{s}), a framework for embodied agent simulation based on Large Language Models (LLMs) that incorporates multiple psychological theoretical principles.
Using simulation, we expand real counseling case data into a nuanced embodied cognitive memory space and generate dialogue data based on high-frequency counseling questions.
We validated our framework using the D$^4$ \cite{yao-etal-2022-d4} dataset. First, we created a public ECAs dataset through batch simulations based on D$^4$. Licensed counselors evaluated our method, demonstrating that it significantly outperforms baselines in simulation authenticity and necessity. Additionally, two LLM-based automated evaluation methods were employed to confirm the higher quality of the generated dialogues compared to the baselines. 
The source code and dataset are available at https://github.com/AIR-DISCOVER/ECAs-Dataset.
\end{abstract}


\section{Introduction}

Mental health counseling data plays a crucial role in multiple applications: training novice counselors \cite{kuehne2018standardized}, developing AI-assisted counseling systems \cite{furlan2021natural, tang2025ca+}, and automating mental health diagnosis \cite{pingexperience}. However, the highly private and sensitive nature of psychological counseling creates significant barriers to collecting and sharing real counseling dialogue data \cite{miner2019key}, hindering both AI development and professional training in this field.

Current approaches to data scarcity fall into two categories. The first relies on simulated dialogues created by human, which offer high authenticity but are costly and may introduce sampling biases due to limited expert availability and individual perspectives \cite{schatzmann2005quantitative}. The second leverages AI-based data synthesis and augmentation, as demonstrated by domain-specific models like HEAL \cite{yuan-etal-2024-continued}. Although these AI-based approaches provide better scalability, they typically capture only surface-level language patterns rather than the underlying psychological complexity \cite{kjell2024beyond}. This limitation often results in distribution shift and pattern collapse during self-training loops, leading to progressive model degradation \cite{shumailov2024ai}.

Recent advances in LLM-based agent simulation show promising results in generating high-fidelity data for social science research, particularly through embodied agents capable of simulating social experiences and interactions \cite{liang-etal-2024-encouraging,wang-etal-2024-unleashing,wang-etal-2023-humanoid, zheng2024large}. Notable examples include generative agents for social behavior modeling \cite{10.1145/3586183.3606763} and socially aligned language models \cite{liu2024training}.

However, simulating psychological counseling presents unique challenges beyond general social simulation. Mental processes, especially those of clients with psychological conditions, are inherently complex and hidden beneath observable behaviors \cite{sircova2015simulating}. While recent works like patient-Psi \cite{wang-etal-2024-patient} and Roleplay-doh \cite{louie-etal-2024-roleplay} have made initial progress, effective simulation requires deep integration with psychological theories and counseling principles. Therefore, our core research question is: How can we develop LLM-based simulations grounded in psychological and counseling theories to generate authentic, rich counseling dialogue data that facilitates research and development in this field?

To address these challenges and advance the field of psychological counseling simulation, we introduce and implement a novel framework called ECAs designed to simulate the \underline{\textbf{E}}mbodied \underline{\textbf{C}}onvers\-ational \underline{\textbf{A}}gent\underline{\textbf{s}} in psychological counseling, integrating counseling theories of Cognitive Behavioral Therapy (CBT) \cite{beck2021cognitivebehavior} with diagnostic frameworks. Our methodology begins with an extensive review of psychological counseling theories, culminating in the formulation of six essential principles and preliminaries for simulation. Leveraging the capabilities of LLMs, we then develop a sophisticated method to expand real counseling case data, creating a highly realistic and nuanced embodied cognitive memory space. Based on this, we construct counselor and client agents that generate dialogue data simulating interactions between counselors and clients, carefully crafted based on high-frequency counseling questions. To rigorously assess the efficacy of our ECAs framework, we employ the D$^4$ \cite{yao-etal-2022-d4}  dataset as a benchmark and involve licensed human counselors in the evaluation process.

Our key contributions include:
i) We introduce a novel LLM-based social simulation framework for nuanced embodied conversational agents in psychological counseling. ii) By reviewing psychological theories, we derive six simulation principles for the ECAs framework. iii) The quality and authenticity of our simulations are validated through both human expert evaluation and the D$^4$ benchmark.
iv) We generate a public ECAs dataset to support future research, along with two LLM-based automated methods for evaluating dialogue quality.

\section{Principles and Preliminaries for Simulation}

The complexity of psychological counseling simulation necessitates a multi-faceted theoretical foundation, as it requires modeling both intricate human characteristics and professional therapeutic interactions. In this section, we introduce six Simulation Principles and Preliminaries (SPs) that enable ECAs to generate high-fidelity counseling conversations grounded in established psychology and psychotherapy theories to address key simulation challenges.

\textbf{SP1: Represent Comprehensive Life-Stage Experiences.}
The primary principle of designing ECAs that authentically represent a client's life experiences across various stages. This involves creating a comprehensive framework for factual memories that incorporates both subjective experiences and event data. The approach aims to capture key long-term memories as well as recent memories, enhancing the understanding of the client's life trajectory. 
This goal is grounded in embodied cognition theory and research on situational cues in memory recall, which emphasize the importance of comprehensive and authentic representations in psychological counseling

Formally, let $L = {l_1, \dots, l_n}$ be the set of life experiences, each with temporal marker $t_i$ and subjective significance $s_i$. The life-experience representation is $M(c) = f(L, T, S)$, where $T$ and $S$ are the temporal and significance mappings. Experiences are categorized as past or recent; past experiences are further divided by age into Childhood, Adolescence, Youth, and Middle Age. Let $L_r$, $T_r$, and $S_r$ denote recent experiences and their mappings. $M(c)$ is represented:

\begin{equation}
M(c) = \sum_{i=1}^{4} \alpha_i S_i \cdot f(L_i, T_i) + \beta S_{\text{r}} \cdot f(L_{\text{r}}, T_{\text{r}})
\end{equation}
Here, $\alpha$ weights the four past stages, reflecting their stronger lasting impact on the client’s psychological state, while $\beta$ assigns a comparatively lower weight to recent experiences, reflecting their relatively smaller impact. 



\textbf{SP2: Simulate Client's Cognitive Processes.}
Mental disorders manifest through distinct patterns of thinking and behavior~\cite{beck2024cognitive,ellis1962reason}. Counseling theories provide systematic frameworks to understand these patterns, with Cognitive Behavioral Therapy (CBT) being particularly influential in modeling thought structures \cite{beck2020cognitive}.
Formally, let $B_c = {b_1, \dots, b_n}$ be core beliefs, $B_i = {i_1, \dots, i_m}$ intermediate beliefs, and $A = {a_1, \dots, a_k}$ automatic thoughts, jointly shaping thought patterns $P$. The cognitive process is $C(c) = g(B_c, B_i, A, P)$, where $P$ maps beliefs and experiences to automatic thoughts:
\begin{equation}
(b_n, i_m, l_n) \rightarrow a_k
\end{equation}
It models how core beliefs, intermediate beliefs and experiences jointly shape automatic thoughts.

\begin{figure*}[ht]
    \centering
    \includegraphics[width=0.93\textwidth]{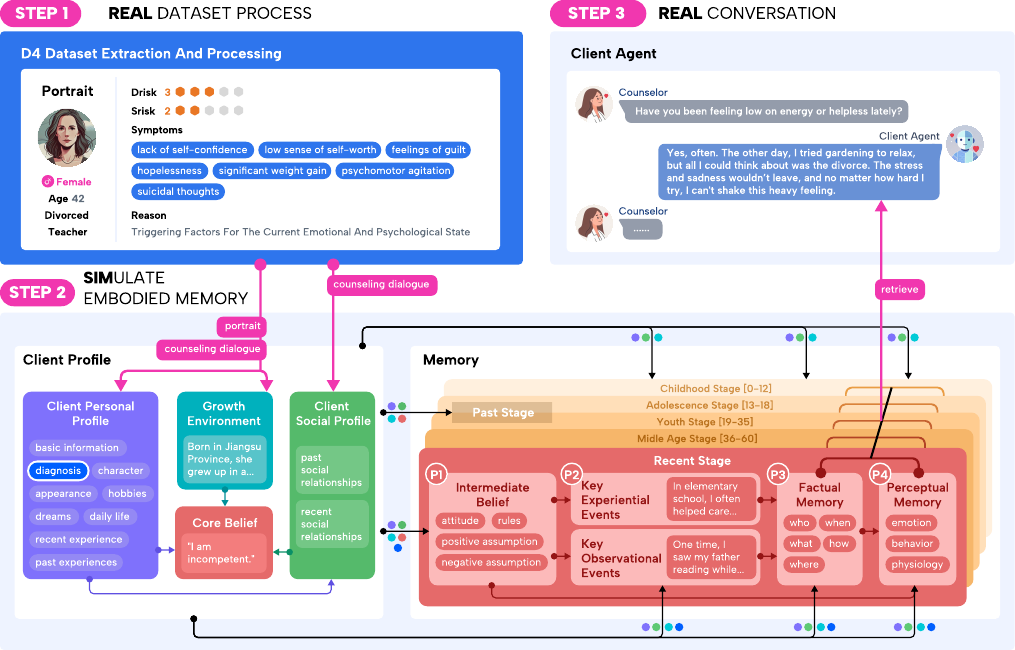}
    \caption{\textbf{ECAs Framework Overview.} The process consists of three steps: \textbf{Step 1}, extracting base information for the \textit{Client Agent} from real datasets; \textbf{Step 2}, expanding the agent's two profiles during memory simulation to form a complete Client Profile, and generating a embodied memory space, including beliefs, cognitive processes, and memories, based on this comprehensive profile; \textbf{Step 3}, dynamically retrieving context-relevant memories during real conversations to ensure realism and consistency.}
    \label{fig:framwork}
\end{figure*}

\textbf{SP3: Integrate Detailed Perceptual Memories.}
Subjective emotions are crucial in psychological counseling. Ianì notes that events consist of perceptual information, reactivating sensorimotor circuits during recall \cite{iani2019embodied}. Culbertson highlights `deep memory' as bodily recollection of trauma, revealing responses beyond verbal expression \cite{culbertson1995embodied}. 
We therefore integrate perceptual memories of clients’ subjective experiences. This enhanced emotional authenticity will enable more effective and true-to-life counseling simulations.  
Let $V(c) = h(\xi_e, \xi_b, \xi_p, R)$ denote perceptual memory, where $\xi_e, \xi_b, \xi_p$ are emotional, behavioral, and physiological responses. $R$ is represented as:
\begin{equation}
(\sum_{n=1}^{n} b_n, \sum_{m=1}^{m} i_m, \sum_{k=1}^{k} a_k) \rightarrow \xi_e, \xi_b, \xi_p
\end{equation}
It models how core beliefs, intermediate beliefs and automatic thoughts jointly shape these responses.



\textbf{SP4: Model Social Interactions and Relationships.}
The principle is to simulate the complex web of social interactions and relationships that shape a client's experiences and behaviors. This framework aims to capture the nuances of interpersonal dynamics, including familial bonds, friendships, professional relationships, and broader social connections. By incorporating theories of social psychology and attachment \cite{bowlby1980attachmentandloss}, the goal is to create a realistic representation of how social relationships influence a client's thoughts, emotions, and actions. 
Let $R(c) = {r_1, \dots, r_n}$ denote social relationships across five developmental stages. Each $r_i$ encodes relationship networks and interaction patterns for a developmental period, capturing the evolution of social connections and their influences. $R(c)$ is defined as:
\begin{equation}
R(c) = \sum_{i=1}^{5} w_i \left( {Density}(r_i) + {Familiarity}(r_i) \right)
\end{equation}
Where $w_i$ are coefficients that weigh the importance of relational Density and Familiarity across different stages.

\textbf{SP5: Maintain Consistency in Data Synthesis.}
This principle aims to preserve accuracy and coherence in client portraits by aligning simulated social environments, cultural backgrounds, and behavioral traits with real-world client profiles. We seek to ensure temporal consistency in the sequence of simulated events and memories, while maintaining contextual accuracy in generated physical environments and social interactions. 
A key objective is to address limitations of LLM-based data augmentation in complex and sensitive counseling scenarios, as highlighted by recent research \cite{fei-etal-2023-mitigating,10.1145/3597307,cilliers2020wearable}, 
aiming to mitigate issues of data bias and inconsistency often encountered in sophisticated NLP tasks, particularly in sensitive domains like psychological counseling. 

\textbf{PS6: Enable Context-Driven Memory Retrieval.}
\label{sec:DG6}
AI agents for simulated clients require context-driven memory retrieval to provide realistic, adaptive responses. Systems like Patient-Psi \cite{wang-etal-2024-patient} and Roleplay-doh \cite{louie-etal-2024-roleplay} demonstrate this need. Effective memory retrieval enables AI agents to provide authentic interactions in healthcare training \cite{li2024leveraginglargelanguagemodel}. Barsalou's work demonstrates that situational triggers activate embodied experiences through pattern completion and inference \cite{barsalou2008grounded}, enhancing contextual relevance and memory retrieval in simulated counseling sessions.

We incorporate a context-aware dynamic memory retrieval mechanism in our ECA framework. It can automatically access related memories, enhancing dialogue depth and authenticity in simulated client interactions. For counseling questions from the D$^4$ dataset, at dialogue time step $t$ the retrieval function $m_t = f(q_t, M(c), H_t)$ selects relevant memories from the client memory space $M(c)$ based on the counselor’s question $q_t$ and dialogue history $H_t$, where $m_t$ guides the generation of high-quality responses.

\section{ECAs Framework}


This section introduces our ECAs framework for generating psychological counseling interactions with embodied \textit{Client Agent}s, covering memory space construction and  high-quality dialogues generation. Section~\ref{r_d_p} extracts base information in real client data from D$^4$ dataset. Section~\ref{s_e_m} simulates personal profile, social profile, and embodied memory. Section~\ref{r_c} describes memory retrieval for realistic counseling. Section~\ref{p_s_d_i} reports a pilot study and design iteration based on therapist feedback to refine memory depth and relevance.

\subsection{\textbf{REAL} Dataset Process}
\label{r_d_p}
To ensure that the generated embodied memories are grounded in credible real-life data and are close to the memories of clients with psychological problems
[SP5], we selected the D$^4$ Chinese dialogue dataset, which includes real client data and depression diagnosis information, as the key basis for data generation. The portrait of the client collected during the Portrait Collection phase of D$^4$ is used as the foundational information for the \textit{Client Agent}. The real counseling dialogue data between the client and counselor is extracted to provide context for profile generation, and the summary from the Professional Diagnosis serves as the description of the \textit{Client Agent}'s current status.

\subsection{\textbf{SIM}ulate Embodied Memory}
\label{s_e_m}
\subsubsection{Client Profile Generation}
\begin{figure}[ht]
    \centering
    \includegraphics[width=0.91\linewidth]{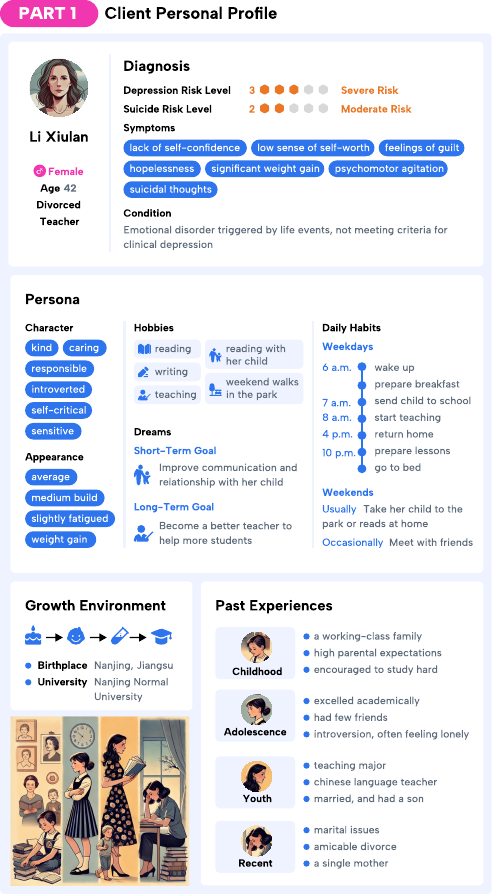}
    \caption{\textbf{Client Personal Profile.} 
    The persona describes the \textit{Client Agent}'s basic information such as name, personality, and appearance, along with background and past experiences to form a complete psychological trajectory.}
    \label{fig:personal profile}
\end{figure}

\begin{figure}[ht]
    \centering
    \includegraphics[width=1\linewidth]{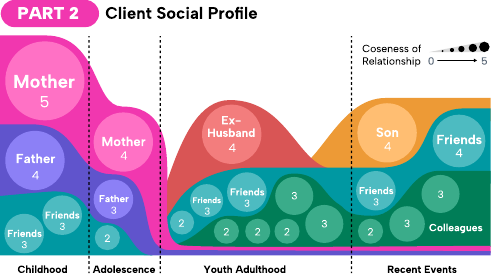}
    \caption{\textbf{Client Social Profile.} The evolution of the \textit{Client Agent}'s social network and relationships over time is reflected, reinforcing the consistency between the social interaction memories and personal profile.}
    \label{fig:social profile}
\end{figure}

Using real data extracted from D$^4$, the \textit{Client Agent}'s profile is generated in two parts: Client Personal Profile (see Figure~\ref{fig:personal profile}) and Client Social Profile (see Figure~\ref{fig:social profile}), ensuring a high degree of consistency between the agent’s persona and social relationships [SP1, SP4, SP5].

\textbf{Part 1}, ECAs utilize the extracted portrait of the client and counseling dialogue as inputs to generate the Client Personal Profile via LLMs. This profile encompasses the \textit{Client Agent}'s recent stage, personality, appearance, hobbies, dreams, daily habits, and recent experience $L_{recent}$. Additionally, it backtracks the client's growth environment and summarizes past experiences $L_{past}$, constructing a coherent psychological trajectory, and aligning personality, behavior, and emotional states with life history for accurate simulations of client’s emotions and behaviors.
\textbf{Part 2}, to align with Part 1, ECAs simulate the \textit{Client Agent}'s social networks $R(c)$ across past and recent stages based on the Client Personal Profile and counseling dialogue.
It reflects changes in the number and familiarity of social connections over time, providing external insights into how social environments trigger and sustain depressive symptoms.

\subsubsection{Client Embodied Memory Generation}

To address the lack of task-specific information in data generated by LLMs, we designed a 4-phase LLM-based generation paradigm (see Figure~\ref{fig:framwork}) grounded in CBT theory. This paradigm generates the \textit{Client Agent}'s core beliefs $B_c$, intermediate beliefs $B_i$, factual memories from experiences $L$, automatic thoughts $A$ and perceptual memories\footnote{The concepts of core beliefs, intermediate beliefs, and automatic thoughts are derived CBT theory. Factual memories and perceptual memories are novel concepts introduced in this framework to enhance the simulation of client experiences.} $V(c)$ across different life stages based on the two profiles [SP2, SP3, SP5].

\textbf{Phase 1:} Generating beliefs. Using Client Personal Profile and Client Social Profile, we first generate recent-stage core beliefs $B_c$ and intermediate beliefs $B_i$, with $B_c$ guiding memory simulation. Then, by backtracking through descriptions of past stages, environments, and the Client Social Profile, we derive $B_i$ for each past stage. Each intermediate belief $i_m$ within $B_i$ covers attitudes toward the self, others, and the world, the rules the client follows in specific problem areas, and associated positive and negative assumptions.

\textbf{Phase 2:} Identifying key events. The generation process is consistent across both recent and past stages, as well as in Phase 3 and Phase 4, with all stages using the Client Profile as input. Notably, `diagnosis' influences only the recent stage, as illustrated in Algorithm~\ref{alg:key events}. The LLMs analyze key experiential and observational events, which form the basis of the \textit{Client Agent}'s experiences $L$ and cause shifts or cognitive changes at each stage with a timeframe of occurrence. Each event is refined into 1-3 descriptions, establishing cause-effect relationships among them.

\renewcommand{\algorithmicrequire}{\textbf{Input:}}  
\renewcommand{\algorithmicensure}{\textbf{Output:}} 

\begin{algorithm}[ht]
\caption{Event and Memory Generation Across Stages}
\label{alg:key events}
\begin{algorithmic}[1]
    \REQUIRE
    $P$: client personal profile without diagnosis;
    $P^*$: client personal profile with diagnosis;
    $G$: growth environment;
    $S$: client social profile;
    $B$: core and intermediate beliefs;
    List = [1, 2, 3, 4, 5] \COMMENT{1: Childhood, 2: Adolescence, 3: Youth, 4: Middle Age, 5: Recent Stage}
    \ENSURE $E$: Can be key events, factual memories, and perceptual memories
    
    \STATE $Num \gets \text{calculate\_number\_of\_past\_stages}(client.age)$
    
    \FOR{each $i \in \text{List}$}
        \IF{$i \leq Num$}
            \STATE $E \gets P[i] + G + S[i] + B[i]$
        \ELSE
            \STATE $i \gets 5$  
            \COMMENT{Force $i$ to be 5 for the recent stage}
            \STATE $E \gets P^*[i] + G + S[i] + B[i]$
        \ENDIF
    \ENDFOR
\end{algorithmic}
\end{algorithm}

\textbf{Phase 3:} Forming factual memories. Both recent and past key events from Phase 2 are reviewed and optimized for realism. The focus is on enriching factual details and ensuring emotional authenticity while avoiding abstract descriptions. Each key event follows the 4W1H  (Who, What, When, Where, How) format and integrates into a complete event description, which serves as \textit{Client Agent}'s factual memory.

\textbf{Phase 4:} Developing perceptual memories. Automatic thoughts $A$, being rapid responses to situations, stem directly from $B_c$ and $B_i$. Each automatic thought $a_k$ links its corresponding core belief $b_n$ and $i_m$. These thoughts and beliefs are further analyzed to generate likely emotional responses $\xi_e$, behavioral responses $\xi_b$, and physiological responses $\xi_p$, forming the \textit{Client Agent}'s perceptual memory $V(c)$.

\subsection{\textbf{REAL} Conversation}
\label{r_c}

To simulate realistic counseling interactions, the \textit{Client Agent} employs a scenario-driven dynamic retrieval mechanism to extract memories most relevant to the dialogue history $H_t$ and counselor's questions $q_t$ before responding [PS6]. As real clients do not explicitly and actively mention general $B_c$ or $B_i$ during counseling, the retrieved memory types are limited to factual memory from $L$, $V(c)$, and A.

First, LLMs analyze the $H_t$ between the \textit{Client Agent} and the counselor, determine the required memory type for the response, and memories containing matching keywords \( kw \) are retrieved from \( M(c) \), and \( m \) is the \textit{memory} entry:
\begin{equation}
m_{matched}= \{ m \in M(c) \mid {kw}(m) \cap {kw}(q_t \cup H_t) \neq \emptyset \}
\end{equation}

Next, cosine similarity is calculated between the vector representations of the conversation context and each memory, selecting the top-3 most relevant memories:
\begin{equation}
m_t = {sort} \left( \left\{ \frac{v_c \cdot v_m}{\|v_c\| \|v_m\|} \mid v_m \in m_{matched} \right\} \right)[:3]
\end{equation}
Here, \( v_c \) is the vector representation of \( c \), and \( v_m \) is the vector representation of each memory \( m \). This process ensures that the most semantically relevant memories are retrieved and ranked for the response.

\subsection{Pilot Study and Iteration}
\label{p_s_d_i}
\subsubsection{Refinement Process} The pilot study underwent three iterations to refine the \textit{Client Agent}'s embodied memory generation and its use in counseling scenarios. In the first iteration, we consulted a therapist to align the memory construction approach with psychological principles. In the second iteration, two therapists evaluated the consistency of the optimized memory scripts with the client profile. However, feedback revealed that script evaluation alone did not fully capture the memory's function in real conversations and could affect assessments of depression and suicide risk, highlighting the need for a more interactive method. In the third iteration, we generated dialogue data using therapist questions to assess the memory’s application in counseling. An LLM extracted highly relevant memory-related questions from the D$^4$ dataset. While the memory was reasonable and aligned with the client profile, it lacked depth in addressing suicide risk, eating behaviors, and probing questions.

\subsubsection{Result} Based on the findings, we refined the depth and completeness of the extracted questions. We carefully reselected questions from the D$^4$ dataset, focusing on five key areas: depression risk, eating behavior, sleep patterns, suicide risk, and social life. These questions were further expanded to explore onset, frequency, intensity, duration, and context, forming a comprehensive set of fourteen questions aligned with real counseling needs.

\section{Evaluation}

\subsection{Dataset Construction.} 

To support expert and automated evaluations, we construct a dataset using the ECAs framework based on D$^4$ real data.

\begin{table}[ht]
\centering
\setlength{\tabcolsep}{1.1pt}  
\renewcommand{\arraystretch}{0.85}
{\small
\begin{tabular}{cccccccccc}
\toprule
\multirow{2.5}{*}{Dataset} & \multirow{2.5}{*}{Group} & \multirow{2.5}{*}{AP} & \multicolumn{4}{c}{Memory} & \multirow{2.5}{*}{LM} & \multirow{2.5}{*}{Avg.MN} \\ \cmidrule{4-7}
 &  &  & FM & PM & CB & IB &  &  &  \\ 
 \midrule
\begin{tabular}[c]{@{}c@{}}D$^4$\\\cite{yao-etal-2022-d4}\end{tabular} & \begin{tabular}[c]{@{}c@{}}Depression\\clients\end{tabular} & 8 & - & - & - & - & - & - \\ 
\midrule
\begin{tabular}[c]{@{}c@{}}CharacterDial\\\cite{zhou-etal-2024-characterglm}\end{tabular} & \begin{tabular}[c]{@{}c@{}}Social\\Characters\end{tabular} & 6 & \ding{52} & - & - & - & - & 1 \\ 
\midrule
\begin{tabular}[c]{@{}c@{}}PATIENT-$\Psi$-CM\\\cite{wang-etal-2024-patient}\end{tabular} & \begin{tabular}[c]{@{}c@{}}Mental\\clients\end{tabular} & 4 & - & \ding{52} & \ding{52} & \ding{52} & - & 1 \\ 
\midrule
\textbf{\begin{tabular}[c]{@{}c@{}}ECAs-dataset\\(ours)\end{tabular}} & \begin{tabular}[c]{@{}c@{}}\textbf{Depression}\\\textbf{clients}\end{tabular} & \textbf{25} & \ding{52} & \ding{52} & \ding{52} & \ding{52} & \ding{52} & \textbf{134.6} \\ 
\bottomrule
\end{tabular}
}
\caption{Comparison with related datasets. AP: Attributes of Profiles, FM: Fact Memory, PM: Perceptual Memory, CB: Core Belief, IB: Immediate Belief, LM: Life-stage experience and Memory retrieval, MN: Memory Nodes.}
\label{tab:related dataset}
\end{table}

It includes two key components: (1) detailed personal profile and social profile for 451 \textit{Client Agent}s, (2) a comprehensive and substantial embodied memory space for 100 of these \textit{Client Agent}s. Each memory space consists of approximately 400 to 1,500 individual memories, which are grouped into memory nodes. Each memory node encapsulates a complete set of factual memories, perceptual memories, and cognitive processes, capturing diverse stages and experiences across the client’s lifetime.



Our dataset was compared with related datasets and demonstrated its distinctive feature of containing a greater quantity and diversity of embodied thinking information (see Table~\ref{tab:related dataset}). Notably, each \textit{Client Agent} features a richer profile with more attributes and a more extensive embodied memory space containing a greater number of memory nodes. In addition to supporting explicit dialogue generation, embodied thinking information can be broadly applied to various other forms of explicit language behavior.

\subsection{Experimental Design and Settings}

To assess the quality of the embodied memory generated by ECAs, we conducted a two-pronged evaluation of dialogues, involving Expert Evaluation by human counseling professionals and automated evaluation utilizing LLM. 
Dialogues were generated through three methods. Control Group 1 and Control Group 2 both used the original, unextended persona data from the D$^4$ dataset, aligned with the Experimental Group's personas before ECAs-specific memory extension:

\textbf{(1) Experimental Group (ECAs):} 
Dialogues generated by GPT-4o, based on the personas and memory settings from our ECAs-dataset, using 14 high-frequency counseling questions, with 3 repeated for consistency assessment.

\textbf{(2) Control Group 1 (GPT-4o):} 
Dialogues generated by GPT-4o with the same questions as the Experimental Group.

\textbf{(3) Control Group 2 (D$^4$):} Original dialogue data collected from human-to-human interactions in the D$^4$ dataset.

For expert evaluation, we recruited five qualified counseling professionals (with certifications, 200+ hours of counseling experience, and advanced degrees in Applied Psychology) to analyze the quality and efficacy of the dialogues. Each professional evaluated the same six randomly selected \textit{Client Agents} with embodied memory space from the ECAs-dataset, assessing necessity, sufficiency, fidelity, and consistency of their responses.
Expert comments were provided during scoring as justification and categorized as positive (P\_) or negative (N\_) based on the same four dimensions.

 \begin{figure}[ht]
    \centering
    \includegraphics[width=0.91\linewidth]{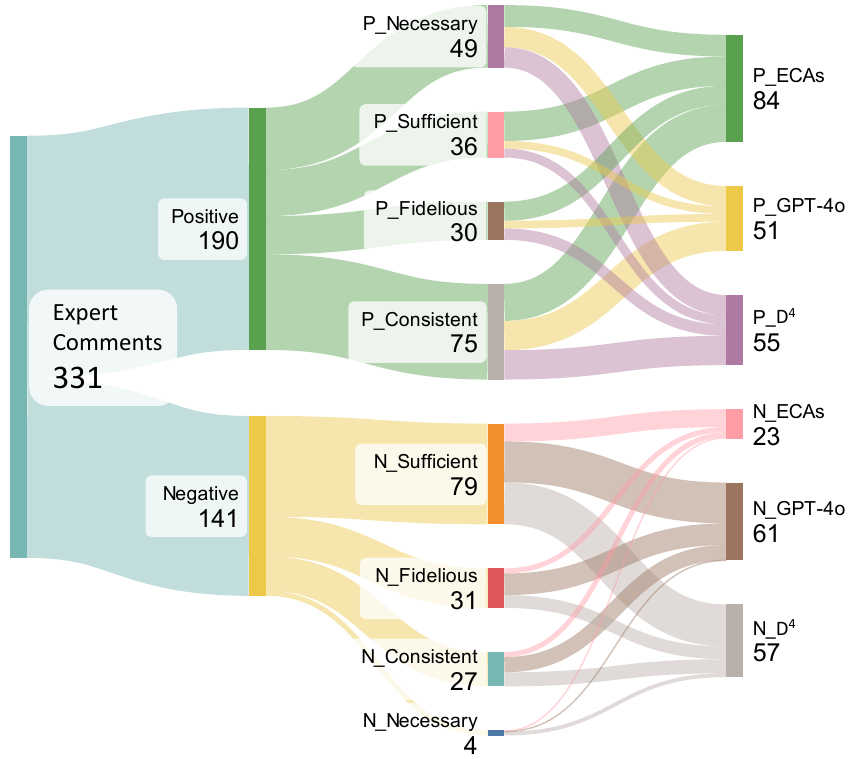}
    \caption{Classification of Positive and Negative Expert Comments. Expert comments are categorized as positive (P\_) or negative (N\_) based on four dimensions, and grouped into ECAs (Ours), GPT-4o, and D$^4$.}
    \label{fig:expert comment}
\end{figure}

Concurrently, we used 100 \textit{Client Agents} with embodied memory space from the ECAs-dataset and implemented two automated evaluation methods using GPT-4o as the evaluator.
The first method assesses dialogue utility in supporting diagnostic decision-making by classifying depression risk and suicide risk into four severity levels (no risk, mild, moderate, and severe). The second method directly assesses dialogue quality based on fidelity, comprehensiveness, consistency, plausibility, and specificity.
This dual approach aimed to offer both nuanced human insights and consistent, large-scale automated analysis, providing a comprehensive evaluation of the embodied memory quality generated by our ECAs framework compared to the baseline approaches.

\subsection{\textbf{Human Expert} Evaluation}

Specifically, experts assessed the necessity and sufficiency of facts and emotional details provided by the client for counseling evaluation, the authenticity of the client's reported experiences and feelings in relation to real clients, and the consistency of the client's responses with their character profile and across similar events.


\begin{figure}[ht]
    \centering
    \includegraphics[width=0.95\linewidth]{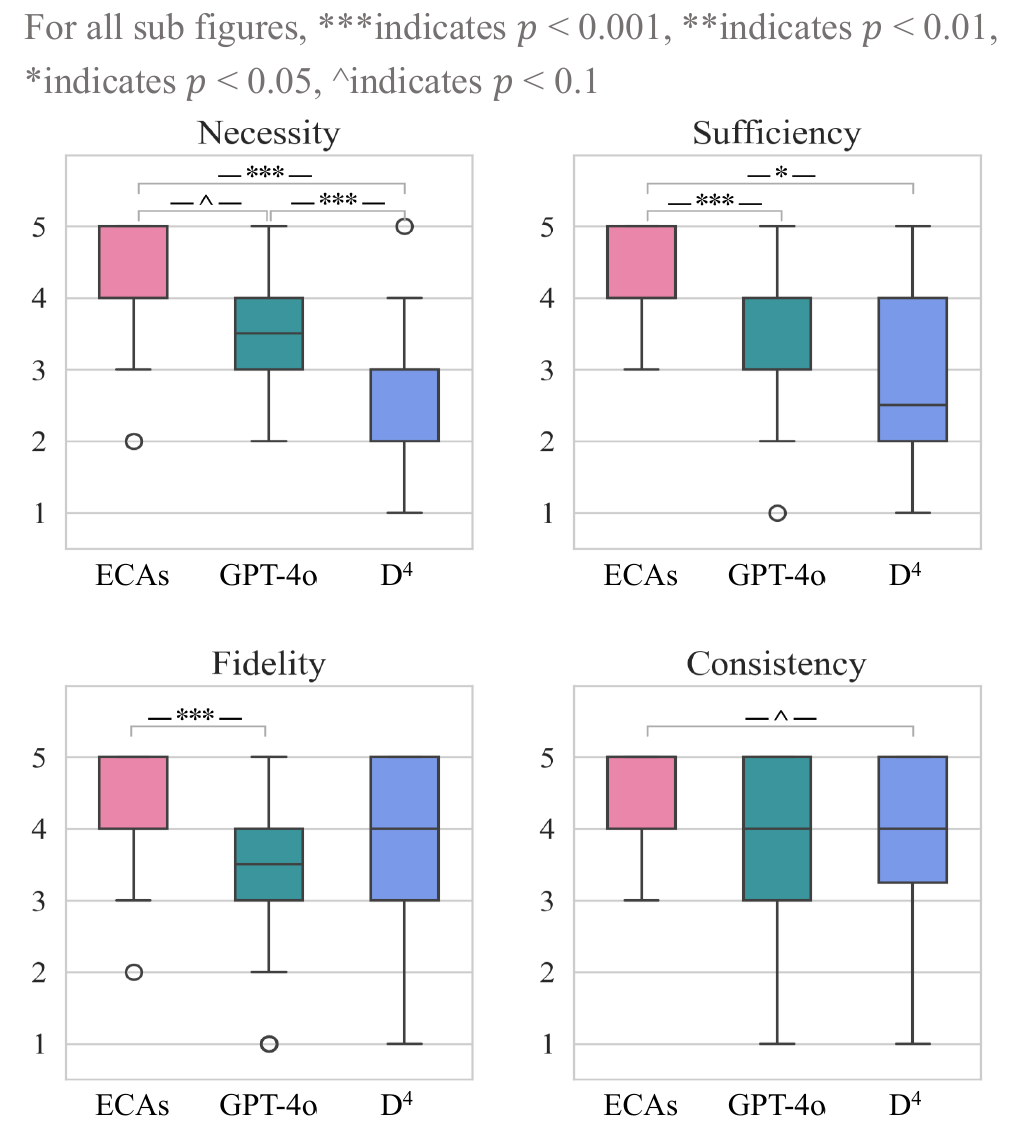}
    \caption{Comparison of ECAs (Ours), GPT-4o, and D$^4$ Performance Across Four Dimensions. Box plots show the distribution of scores for Necessity, Sufficiency, Fidelity, and Consistency. 
    }
    \label{fig:data}
\end{figure}

\subsubsection{Results} As shown in Figure \ref{fig:expert comment}, dialogues generated by ECAs achieved the highest positive and lowest negative expert evaluations, outperforming GPT-4o and D$^4$. Experts regarded the dialogues generated by our method as more sufficient, necessary, fidelious, and consistent. Notably, ECAs received three times more positive and one-third fewer negative evaluations for sufficiency. ECAs dialogues demonstrated stronger relevance to depression diagnosis and significantly higher quality and reliability than the benchmarks.
Figure \ref{fig:data} further highlights the significant improvements demonstrated by our framework across multiple dimensions compared to both GPT-4o and human-generated responses.

In terms of \textbf{Necessity}, a repeated measures ANOVA revealed significant differences among the three conditions, $F(2, 87) = 16.173$, $p < 0.001$, $\eta^2 = 0.271$. Post-hoc tests using Bonferroni correction showed that ECAs marginally outperformed GPT-4o (mean difference $= 0.56667$, $p = 0.074 < 0.1$). ECAs significantly outperformed human responses (mean difference $= 1.40000$, $p < 0.001$), and GPT-4o also significantly outperformed human responses (mean difference $= 0.83333$, $p = 0.003 < 0.01$).
Expert E4 praised the ECAs' performance, noting that \textit{`the additional emotional details provided by ECAs enriched the dialogue, making it more comprehensive'}. Expert E5 emphasized that the ECAs' responses \textit{`contained numerous specific details that \textbf{significantly aid in assessing the client's precise situation and emotional state}'}.

For \textbf{Sufficiency}, the ANOVA again showed significant differences, $F(2, 87) = 15.796$, $p < 0.001$, $\eta^2 = 0.266$. Post-hoc comparisons indicated that ECAs significantly outperformed both GPT-4o (mean difference $= 1.06667$, $p = 0.001$) and human responses (mean difference $= 1.50000$, $p < 0.001$).
Expert E3 noted ECAs' output as \textit{``very much aligns with a client immersed in grief"}, while Expert E5 noted the \textit{`reasonable grief reactions to the loss of a loved one'}. Expert E4 described the responses as \textit{\textbf{`delicate and comprehensive'}}, with Expert E3 further praising the \textit{`vivid examples and nuanced thoughts and emotions'} presented.

Regarding \textbf{Fidelity}, the ANOVA revealed significant differences, $F(2, 87) = 5.188$, $p = 0.007 < 0.01$, $\eta^2 = 0.107$. Post-hoc tests showed that ECAs significantly outperformed GPT-4o (mean difference $= 1.00000$, $p = 0.005 < 0.01$)
Expert E3 highlighted that the ECAs' responses \textit{`consistently addressed the counselor's questions accurately, without deviating from the topic'} and \textit{``\textbf{effectively matched and addressed the questions} at hand"}.

For \textbf{Consistency}, the ANOVA showed significant differences, $F(2, 87) = 3.217$, $p = 0.045 <0.05$, $\eta^2 = 0.069$. There was a marginally significant difference between ECAs and human responses (mean difference $= 0.56667$, $p = 0.079 < 0.1$).
Experts noted that the responses were \textit{'consistent with the client's identity and experiences'} (E5), with E4 adding that \textit{``the client's efforts, lack of self-confidence, concern for parental opinions, and the series of depressive symptoms stemming from graduate school pressure were all highly congruent"} with the expected profile.

These results indicate that the ECAs framework consistently produced higher quality responses across all measured dimensions, with particularly strong improvements in necessity and sufficiency compared to both GPT-4o and human-generated responses.

\subsection{\textbf{Automated Evaluation}}



For the automated evaluation, 
the first automated evaluation focused on classifying depression risk (drisk) and suicide risk (srisk) based on dialogue data. To ensure fairness and mitigate the potential impact of uneven sample distribution on the fairness of results, the four-class classification is evaluated by macro-averaged Precision, Recall, and F1 by sklearn\footnote{\url{https://scikit-learn.org}}. As shown in Table~\ref{tab:risk_result}, dialogues generated by our framework achieve the highest performance across both tasks, demonstrating that embodied memory component provides more vital information and significantly contributes to the diagnostic process.

\begin{table}[ht]
\centering
\begin{tabular}{c|cccc}
\toprule
 Task & Method & Precision & Recall & F1 \\ \midrule
       & GPT-4o & 0.44\small{±0.02} & 0.41\small{±0.05} & 0.35\small{±0.04} \\
       & D$^4$ & 0.49\small{±0.03} & 0.44\small{±0.04} & 0.40\small{±0.05} \\
       \multirow{-3}{*}{drisk} & Ours & \textbf{0.51}\small{±0.04} & \textbf{0.50}\small{±0.08} & \textbf{0.42}\small{±0.04} \\ 
       \midrule
       & GPT-4o & 0.64\small{±0.05} & 0.59\small{±0.03} & 0.59\small{±0.04} \\
       & D$^4$ & 0.67\small{±0.04} & 0.55\small{±0.05} & 0.59\small{±0.05} \\
  \multirow{-3}{*}{srisk} & Ours & \textbf{0.71}\small{±0.08} & \textbf{0.66}\small{±0.05} & \textbf{0.67}\small{±0.06} \\ \bottomrule
\end{tabular}
\caption{Depression and Suicide Severity Classification}
\label{tab:risk_result}
\end{table}

\begin{table}[ht]
    \centering
    \setlength{\tabcolsep}{4pt}  
    \renewcommand{\arraystretch}{1.2} 
    {\small
        \begin{tabular}{c|c c c c c|c}
        \toprule
        Method & F & C1 & C2 & P & S & Total \\ 
        \midrule
        D$^4$ & 2.40/5 & 2.91/7 & 1.03/2 & 1.98/3 & 0.96/2 & 9.28/20 \\
        \midrule
        Ours & \textbf{4.66}/5 & \textbf{6.26}/7 & \textbf{1.99}/2 & \textbf{2.98}/3 & \textbf{1.97}/2 & \textbf{17.90}/20 \\ 
        \bottomrule
        \end{tabular}
    }
    \caption{Auto Qulity Evaluation Across Five Dimensions}
    \label{tab:auto evaluation}
\end{table}



For the second automated evaluation, we compared the dialogue quality generated by our method with that from human role-playing in Control Group 2.
Evaluation covered five key metrics: fidelity (F), comprehensiveness (C1), consistency (C2), plausibility (P), and specificity (S), with weighted scores reflecting each metric's importance. 
Results in Tabl~\ref{tab:auto evaluation} show that ECAs-generated dialogues performed strongly across all metrics, especially in fidelity and comprehensiveness, significantly outperforming human-generated dialogues in D$^4$. From the LLM perspective, this indicates that ECAs-generated dialogues provide deep, authentic emotional responses across diverse life experiences while maintaining high consistency and plausibility.

\section{Conclusion}

This paper presents ECAs, a novel framework for simulating embodied client agents in psychological counseling that generates high-fidelity counseling conversational data. Integrating counseling theories By integrating counseling theories with LLMs, we expand real counseling data into a nuanced cognitive memory space to generate realistic dialogues. 
Guided by six principles for simulation derived from counseling theories, our framework was validated using the D$^4$ dataset and evaluated by licensed counselors, demonstrating significant improvements in simulation authenticity and necessity. 
Additionally, automated evaluation methods validated the higher-quality dialogues generated by our framework.
Looking ahead, the ECAs holds immense potential for counselors to generate custom scenarios on-demand, enhancing both training and research capabilities.

\section{Acknowledgments}

This work was supported by the Beijing Municipal Science and Technology Project (Nos. Z231100010323005) and China Natural Science Foundation Youth Fund 62202267.



\bibliography{main}

\end{document}